\documentclass[10pt]{article}
\usepackage[utf8]{inputenc}
\usepackage[T1]{fontenc}
\providecommand{\keywords}[1]{\noindent\textbf{Keywords:} #1}
\usepackage{subcaption}
\usepackage{xcolor}
\usepackage{soul}
\usepackage{graphicx}
\usepackage{booktabs}
\usepackage{rotating}
\usepackage[numbers,sort&compress]{natbib}
\usepackage{newunicodechar}
\newunicodechar{−}{-}
\usepackage{enumitem}
\usepackage{float}
\usepackage{amsmath,amssymb,amsfonts}
\usepackage{textcomp}
\usepackage{url}
\usepackage{algorithm} 
\usepackage{algpseudocode}
\usepackage{threeparttable}
\usepackage{multirow}
\usepackage{array}
\usepackage{makecell}
\usepackage{balance}
\usepackage{ragged2e} 
\usepackage{tabularx}
\usepackage{longtable}
\usepackage{threeparttablex}
\usepackage[left=2.5cm,top=2.5cm,right=2.5cm,bottom=2.5cm,bindingoffset=1.0cm]{geometry}
\usepackage{authblk}
\usepackage{hyperref}
\usepackage{orcidlink}
\usepackage{cleveref}      

\algrenewcommand\algorithmicrequire{\textbf{Input:}}
\algrenewcommand\algorithmicensure{\textbf{Output:}}

\title{Re-evaluating Heart Rate Variability Biomarkers for Glucose Sensing: The Impact of Age Normalisation and Subject-Independent Validation}

\date{}
\author{Md Basit Azam\textsuperscript{1,*}\,\orcidlink{0009-0006-6538-5644}, Sarangthem Ibotombi Singh\textsuperscript{1}\,\orcidlink{0009-0009-2789-0078} (\href{mailto:sis@tezu.ernet.in}{sis@tezu.ernet.in})\\
\small \textsuperscript{1}Department of Computer Science \& Engineering, School of Engineering, Tezpur University, Napaam - 784028, Tezpur, Assam, India.\\
\small *Corresponding Author: Md Basit Azam; \href{mailto:mdbasit@tezu.ernet.in}{mdbasit@tezu.ernet.in}
}

\begin{document}
\flushbottom
\maketitle
\vspace{-2em}
\thispagestyle{empty}

\begin{abstract}
\noindent\textbf{Background:} Heart rate variability (HRV) derived from electrocardiogram (ECG) signals offers a promising non-invasive window into glycemic status; however, existing studies frequently combine distinct glucose measurements and employ validation strategies susceptible to data leakage. Because HRV declines by approximately 3--5\% per decade due to age-related autonomic degeneration, absolute HRV values conflate the effects of aging with diabetes-specific autonomic dysfunction. We hypothesised that normalising HRV features using an age-dependent scaling factor would isolate the diabetes-specific component and improve glycemic status estimation. \\[0.4em]
\textbf{Methods:} We analysed ECG-derived features from 43 male type 2 diabetes patients with strictly separated glycated hemoglobin (HbA1c; n = 29; 3-month glycemic average) and fasting blood glucose (FBG; n = 38; acute status). Leave-one-subject-out (LOSO) cross-validation (CV) with within-fold feature selection and standardisation prevented information leakage. Twenty machine learning algorithms and six age-adjustment methods were compared, with normalisation sensitivity tested across 20 parameter combinations. Statistical validation employed permutation testing (n = 500) and bootstrap 95\% confidence intervals. \\[0.4em]
\textbf{Results:} Extra trees regression achieved the best performance: $R^2 = 0.222$ ($r = 0.476$, $p = 0.009$) for HbA1c and $R^2 = 0.086$ ($r = 0.344$, $p = 0.034$) for FBG, corresponding to mean absolute errors of 1.18 percentage points and 2.27 mmol/L respectively. Permutation testing confirmed that both associations exceeded the chance level ($p = 0.002$). Contrary to our hypothesis, none of the six age-adjustment methods nor any of the 20 sensitivity parameter combinations improved performance, indicating that age-related HRV decline did not confound glycemic estimation in this cohort. CV hygiene differentially affected model families: tree-based ensembles maintained positive performance, whereas linear models collapsed to negative $R^2$ values, revealing substantial bias from conventional practices. Neural networks with minimally configured hyperparameters failed for these sample sizes ($R^2$ ranging from $-8.2$ to $-10{,}879$). \\[0.4em]
\textbf{Conclusions:} Strict within-fold preprocessing fundamentally alters conclusions in HRV-based glycemic status estimation, exposing inflated performance to conventional CV practices. Bootstrap confidence intervals excluding zero (HbA1c $R^2$: [0.13, 0.82]; FBG $R^2$: [0.10, 0.72]) provided statistical evidence for genuine HRV--glycemic associations, but performance remained insufficient for standalone clinical use. This study establishes methodological standards for separating glycemic targets, subject-independent validation with within-fold preprocessing, and comprehensive baselines to advance non-invasive glycemic monitoring research.
\end{abstract}

\vspace{0.5em}

\vspace{0.5em}
\noindent\textbf{Clinical Trial Number:} Not Applicable

\section{Introduction}
\label{sec1}
\subsection{The Clinical Need for Non-Invasive Glucose Monitoring}

Diabetes mellitus is one of the most significant global health challenges of the 21st century, with approximately 589 million adults (aged 20--79 years) living with diabetes and is projected to reach 853 million by 2050 \cite{facts_figures}. It remains a leading contributor to cardiovascular disease, nephropathy, retinopathy, and premature mortality, with global healthcare expenditures exceeding \$966 billion annually \cite{hossain_diabetes_2024}. Effective glucose monitoring constitutes the cornerstone of diabetes management, enabling timely therapeutic adjustments, preventing acute complications, and optimising long-term glycemic control \cite{american_diabetes_association_professional_practice_committee_2_2022}.

Current glucose monitoring paradigms rely predominantly on invasive methods. Self-monitoring of blood glucose requires repeated finger-prick capillary blood sampling, whereas continuous glucose monitoring (CGM) systems necessitate subcutaneous sensor insertion with replacement every 7-14 days \cite{danne_international_2017}. Although CGM technology has revolutionized diabetes management by providing real-time glucose trends and alerts, significant barriers persist, including sensor costs ranging from \$100-400 per month, insertion site reactions, calibration requirements, and the psychological burden of wearing medical devices continuously \cite{polonsky_what_2014}. These factors contribute to suboptimal monitoring adherence, particularly in resource-limited settings and among populations facing diabetes-related distress \cite{fisher_diabetes_2016}. The clinical imperative for truly non-invasive glucose monitoring has driven extensive research into optical, electromagnetic, and biosignal-based approaches \cite{shang_products_2022}. Physiological signals derived from electrocardiography (ECG) have attracted considerable attention because of the ubiquity of cardiac monitoring devices, availability of consumer-grade wearables, and established relationship between autonomic function and metabolic regulation \cite{porumb_precision_2020, elvebakk_evaluation_2018}.  

Heart rate variability (HRV) analysis derived from ECG has several benefits over other non-invasive techniques, such as photoplethysmography (PPG), for the estimation of the glycemic state. Although PPG can record pulse rate variability (PRV) as a peripheral indicator of HRV, ECG allows the direct evaluation of cardiac electrical activity with high temporal resolution for precise R-peak detection. Although PRV and HRV show a high correlation in resting states, PPG signals are prone to artifacts from motion, vasoconstriction, and skin pigmentation, which can reduce the accuracy of PRV, and discrepancies between PRV and HRV tend to be higher during physiological stress and in clinical settings \cite{schafer_how_2013}. In fact, some researchers have proposed that PRV should be considered a separate biomarker from HRV due to the effects of pulse transit time variability and peripheral vascular dynamics on PPG signal-derived parameters \cite{mejia-mejia_heart_2020}. Moreover, ECG allows morphological analysis (QT intervals and ST segments) that convey metabolic information beyond autonomic regulation; QTc interval prolongation during hypoglycemia and ST-segment changes with glycemic variations are established electrophysiological markers of metabolic disturbances \cite{stern_ecg_2009}. However, as this study demonstrates, applying strict subject-independent validation with within-fold preprocessing reveals that the performance is substantially lower and more realistic than conventional practices suggest.

\subsection{Physiological Basis: Autonomic Nervous System and Glucose Homeostasis}

The rationale for ECG-based glucose estimation relies on the bidirectional relationship between glycemic status and autonomic nervous system (ANS) function. The ANS, comprising sympathetic and parasympathetic divisions, plays a central role in glucose homeostasis through multiple mechanisms \cite{vinik_autonomic_2011}. Sympathetic activation stimulates hepatic gluconeogenesis and glycogenolysis, inhibits insulin secretion, and promotes glucagon release, collectively increasing blood glucose levels \cite{nonogaki_new_2000}. Conversely, parasympathetic activity via the vagus nerve stimulates insulin secretion from pancreatic beta cells and facilitates glucose uptake in the peripheral tissues \cite{thorens_neural_2014}.

Heart rate variability (HRV), defined as the variation in time intervals between consecutive heartbeats, provides a non-invasive window for ANS function \cite{noauthor_heart_1996}. The Task Force of the European Society of Cardiology and the North American Society of Pacing and Electrophysiology established standardized HRV metrics in 1996, distinguishing time-domain measures (e.g., standard deviation of NN intervals [SDNN], root mean square of successive differences [RMSSD], percentage of successive intervals differing by $>$50 ms [pNN50]) from frequency-domain components (e.g., low-frequency [LF] and high-frequency [HF] power) \cite{noauthor_heart_1996}. Time-domain metrics, particularly RMSSD and pNN50, predominantly reflect parasympathetic (vagal) modulation, whereas frequency-domain LF power represents a mixture of sympathetic and parasympathetic influences \cite{noauthor_heart_1996}.

Multiple lines of evidence support the HRV-glucose relationship. Rothberg et al. (2016) demonstrated significant correlations between HRV parameters and blood glucose levels in both diabetic and non-diabetic populations, with frequency-domain measures showing particular sensitivity to glycemic fluctuations \cite{rothberg_association_2016}. Kajisa et al. (2024) reported moderate negative correlations ($r \approx -0.45$) between nocturnal glucose levels and HRV in healthy adults during sleep \cite{kajisa_correlation_2024}. Im et al. (2023) found that poor glycemic control is associated with significantly reduced HRV, whereas well-controlled glucose profiles correlate with preserved autonomic function \cite{im_real-time_2023}. Klimontov et al. (2016) demonstrated that HRV metrics correlated with interstitial glucose variability in insulin-treated women with type 2 diabetes \cite{klimontov_heart_2016}.

These associations reflect the underlying pathophysiology: chronic hyperglycemia induces oxidative stress and advanced glycation end-product accumulation, damaging autonomic nerve fibers and leading to cardiac autonomic neuropathy (CAN) \cite{pop-busui_cardiac_2010}. CAN affects up to 60\% of patients with longstanding diabetes and manifests clinically as reduced HRV, resting tachycardia, and impaired heart rate response to physiological stimuli \cite{spallone_cardiovascular_2011}. This pathophysiological linkage provides a mechanistic rationale for using HRV as a potential biomarker of glycemic status.

\subsection{Age-Related Autonomic Decline and the Normalization Challenge}

Age-related autonomic decline is a critical confounding factor in HRV-based glycemic status estimation. Longitudinal studies have demonstrated progressive reductions in HRV across the adult lifespan, with SDNN decreasing by approximately 3-5\% per decade \cite{bonnemeier_circadian_2003}. Umetani et al. (1998) documented that 24-hour HRV metrics decline linearly with age, with the most pronounced reductions occurring in parasympathetic-mediated components \cite{umetani_twenty-four_1998}. Zhang et al. (2007) confirmed these findings, demonstrating that age accounts for 10-20\% of HRV variance in healthy populations \cite{zhang_effect_2007}.

This age-related decline poses substantial challenges for the development of HRV-based biomarkers. A 70-year-old patient with excellent glycemic control may exhibit lower absolute HRV values than a 40-year-old patient with poor glycemic control, potentially confounding estimation models. The Task Force guidelines acknowledge this limitation but do not prescribe formal age-normalization methods \cite{noauthor_heart_1996}. Several approaches have been proposed in the literature, and Stojmenski et al. (2023) demonstrated that age and sex normalization significantly improved HRV-based glycemic status estimation performance \cite{stojmenski_age_2023}. However, the optimal normalization strategy remains undefined, with options including simple division by age, residualization (regressing out age effects), z-score standardization within the age strata, and polynomial interaction terms.

\subsection{Sleep-Stage Specificity and Autonomic Function}

The recognition that sleep architecture modulates autonomic activity has prompted the investigation of sleep-stage-specific HRV for glycemic status estimation. Sleep comprises distinct physiological states with characteristic autonomic profiles: non-rapid eye movement (NREM) sleep features progressive parasympathetic dominance, whereas rapid eye movement (REM) sleep exhibits sympathetic surges and increased HRV \cite{brandenberger_age-related_2003}. Deep sleep (NREM stage N3) represents the period of maximal parasympathetic tone and metabolic quiescence, whereas REM sleep demonstrates variable autonomic activity that resembles wakefulness \cite{somers_sympathetic-nerve_1993}.

Cheng et al. (2023) demonstrated that HRV metrics captured during stable and REM sleep stages were significantly correlated with fasting glucose and HbA1c levels in patients with type 2 diabetes \cite{cheng_heart_2023}. Martyn-Nemeth et al. (2018) reported that poor sleep quality in patients with type 1 diabetes was associated with greater nocturnal glycemic variability \cite{martyn-nemeth_poor_2018}. These findings suggest that sleep-stage-stratified HRV analysis may enhance predictive modeling by capturing autonomic dynamics under controlled physiological conditions. Radha et al. (2019) further demonstrated that HRV features alone can achieve accurate sleep stage classification, enabling sleep-aware analysis without polysomnography \cite{radha_sleep_2019}.

\subsection{Methodological Concerns in Existing Literature}

Despite promising preliminary findings, the HRV-glycemic status estimation literature suffers from significant methodological heterogeneity that limits interpretability and clinical translation. Three principal concerns have emerged from the critical appraisal of existing studies.

First, we confirmed the fundamental difference in glucose measurements. Studies frequently combine or interchange HbA1c (reflecting 3-month average glycemic control through erythrocyte glycation \cite{american_diabetes_association_professional_practice_committee_2_2022}) with fasting blood glucose (FBG, representing acute metabolic status) or CGM-derived metrics (capturing real-time glucose dynamics). These measurements have different physiological meanings, temporal scales, and clinical interpretations. HbA1c correlates with chronic metabolic exposure and the risk of complications, whereas FBG reflects overnight hepatic glucose output and acute insulin sensitivity \cite{american_diabetes_association_professional_practice_committee_2_2022}. Combining these targets confounds model interpretation and limits mechanistic insight.

Second, the validation strategies are susceptible to information leakage. Most published studies employed random k-fold cross-validation (CV), which can leak information when multiple measurements are derived from the same subject \cite{varoquaux_cross-validation_2018, saeb_need_2017}. In physiological datasets, where within-subject correlation exceeds between-subject variation, random splits may assign correlated samples to both training and test sets, artificially inflating performance estimates. Varoquaux (2018) demonstrated that small sample sizes combined with random CV produce large error bars and unreliable performance estimates \cite{varoquaux_cross-validation_2018}. Saeb et al. (2017) emphasised that clinical machine learning requires validation approximating the intended use case, typically generalisation to new, unseen subjects \cite{saeb_need_2017}. Leave-one-subject-out (LOSO) validation addresses this concern by ensuring complete subject-level separation between the training and test data. Another methodological issue, often overlooked, relates to CV hygiene, namely whether feature selection and scaling are performed inside each CV loop or on the entire data set before splitting. If feature selection is trained on all subjects, including those assigned to the test set, the selected features will have access to the target information of the test subjects, which is a subtle but important form of data leakage \cite{varma_bias_2006}. This study addresses these issues by performing all preprocessing steps inside each LOSO loop.

This study adds to the HRV-glucose body of the literature as a validation-focused, negative-result study. Rather than claiming predictive paradigm shifts, we showed that by following strict methodological principles using distinct glycemic goals, subject-agnostic LOSO validation with fold-level preprocessing, and thorough comparisons of baselines across twenty methods the performance is substantially lower and more realistic than usual, thus setting the record straight for the community. The findings are both positive (statistically significant associations that pass permutation testing) and negative (no advantage of age normalization, catastrophic performance of neural networks with standard parameters and small sample sizes), as expected in best practices for open biomedical informatics research.

\subsection{Distinguishing CGM from Spot Glucose Measurements}

A critical distinction that warrants explicit clarification is the nature of the glucose data in this study. CGM provides interstitial glucose measurements every 1-15 minutes, enabling the temporal analysis of glucose-HRV relationships and the prediction of future glucose values \cite{danne_international_2017}. In contrast, spot glucose measurements, including laboratory HbA1c and FBG levels, represent single-timepoint assessments without continuous temporal correspondence to ECG recordings.

The present study utilised spot glucose measurements (HbA1c and FBG) obtained during hospitalisation, but not CGM data. This distinction has profound implications for interpretation: we analysed cross-sectional associations between ECG-derived features and glycemic status across subjects, rather than within-subject temporal prediction. The ECG-glucose relationship examined here is correlational and cross-sectional, and not predictive of real-time glucose fluctuations. This constraint reflects dataset availability but also represents a more conservative analytical approach that is less susceptible to overfitting on within-subject temporal autocorrelation.

\subsection{Study Objectives and Hypotheses}

Given the methodological concerns identified above, this study undertook a rigorous reappraisal of ECG-derived feature associations with glycemic status. We address four gaps in the literature:

\begin{enumerate}
    \item We created strictly separate cohorts for HbA1c (3-month glycemic average) and FBG (acute status) levels, recognising their fundamentally different physiological meanings.
    \item We implemented the LOSO CV to prevent information leakage and provide unbiased performance estimates that are essential for clinical translation.
    \item We systematically evaluated 20 machine learning algorithms, including naive baselines, regularised linear models, tree ensembles, support vector machines, and neural networks, along with six age-adjustment methods with a 20-combination sensitivity grid for age normalisation parameters.
    \item We enforced strict CV hygiene, feature selection, and standardisation within each LOSO fold to prevent information leakage, and quantified the impact of this correction on reported performance.
\end{enumerate}

Our primary hypothesis was that age-normalised HRV features would significantly improve glycemic status estimation accuracy by correcting for age-related autonomic decline. Secondary objectives included characterising feature domain contributions (clinical vs. ECG vs. HRV), evaluating neural network performance on small samples, and establishing methodological standards for future research in this domain.

\section{Methodology}

\subsection{Study Design and Participants}

\begin{figure*}
    \centering
    \includegraphics[width=\linewidth]{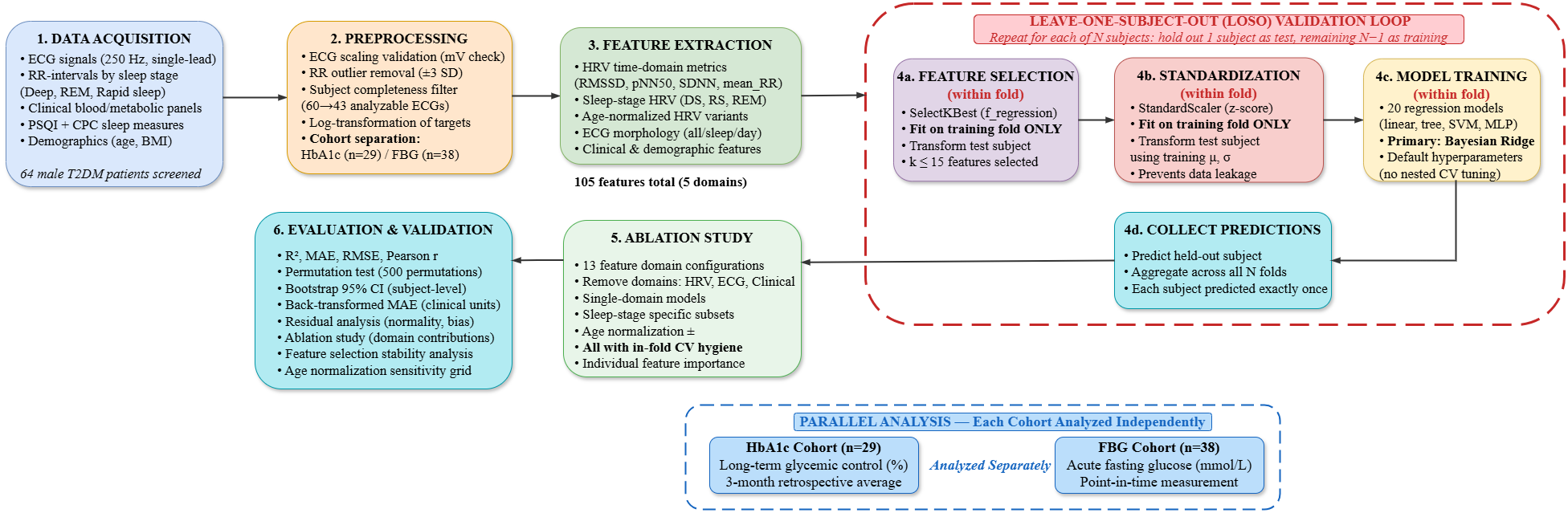} 
    \caption{Overview of the computational pipeline for ECG-based glycemic estimation. HbA1c and FBG cohorts are analyzed separately due to distinct physiological timescales (3-month average vs. acute status)}
    \label{fig:block_daigram}
\end{figure*}

The dataset used in this study comprised a collection of 24-hour ECG signals, ECG analysis results based on circadian rhythm and R-peak detection, results of sleep quality assessment, and clinical indicators of metabolic function acquired from 60 male inpatients with type 2 diabetes mellitus (T2DM) \cite{cheng_dataset_2023}. The study population consisted of male inpatients with T2DM. The initial screening included 64 patients with the following exclusion criteria: female sex, type 1 diabetes mellitus, history of stroke, subacute myocardial infarction, kidney or liver transplant, other systemic disorders, current recreational drug or alcohol abuse, and morbid obesity (body mass index$>$40 kg/m\textsuperscript{2}). Four patients were excluded because they were female, yielding 60 male patients. All participants were hospitalised for diabetes management and of Asian (Chinese) ethnicity. Of these 60 subjects, 43 had analyzable 24-hour ECG recordings following quality control assessment. Seventeen recordings (28.3\%) were excluded because of low signal quality, insufficient recording duration, presence of atrial fibrillation, or severe arrhythmias. The final cohort characteristics were a mean age of $50\pm16$ years (range 28-82 years). This male-only, single-ethnicity cohort represents a significant limitation affecting generalizability, as discussed in Section 4.2. The complete analytical pipeline is illustrated in Figure~\ref{fig:block_daigram}. In summary, after excluding four female patients, 64 were recruited $\rightarrow$ 60 after exclusion of 4 female patients $\rightarrow$ 43 after discarding 17 ECG recordings for quality reasons $\rightarrow$ 40 subjects with complete feature sets after removing 3 with missing RR-interval files or incomplete clinical records. HbA1c measurements were available for 29 of these subjects, and FBG for 38 patients, forming the two analysis cohorts.

\subsection{Target Variable Separation}

We created strictly separate cohorts for the two available glucose measurements, recognising their fundamentally different physiological meanings.

\begin{enumerate}
    \item \textbf{HbA1c Cohort (n=29):} Glycated hemoglobin (\%), reflecting approximately 3-month average glycemic control. Mean $8.50\pm1.68\%$ (range 5.60-11.80
    \item \textbf{FBG Cohort (n=38):} FBG (mmol/L), representing the acute metabolic status measured on admission. Mean $9.16\pm3.07$ mmol/L (range 5.20-16.50 mmol/L). Conversion: 1 mmol/L = 18 mg/dL.
\end{enumerate}

Both targets were log-transformed to address distributional skewness and improve the regression stability \cite{keene_log_1995}. This separation prevents the common error of combining fundamentally different glucose metrics, which confounds the physiological interpretation.

\subsection{ECG Acquisition and Signal Processing}
The 24-hour ECG monitoring began at 22:00 h on the second day of hospitalization using an FDA-approved ambulatory single-lead Holter ECG monitor (DynaDx Corporation, CA, USA; sampling rate = 250 Hz) \cite{cheng_dataset_2023}. Quality assessment of all ECG recordings for noise, artifacts, and ectopic activity was performed by the original researchers, and seventeen recordings were removed because of poor data quality, short recording time, or presence of atrial fibrillation or severe arrhythmia. Sleep stages were determined using an ECG-based cardiopulmonary coupling (CPC) analysis, which assigned epochs into stable sleep, unstable sleep, and REM sleep. The dataset offered pre-detected R-peak to R-peak (RR) intervals corresponding to sleep stages \cite{cheng_dataset_2023}. In our study, we added RR interval quality control steps for feature extraction in HRV: RR intervals labeled as outliers, beyond $\pm 3$ standard deviations from the mean, were removed, and at least 10 intervals per sleep stage were required for HRV feature extraction. ECG signal amplitude validation was performed against biologically plausible values ($\pm 0.1$ to $\pm 10$ mV). 

\subsection{Sleep Stage Classification}
Sleep staging utilised CPC analysis applied to ECG recordings during sleep periods identified from participant-completed sleep logs \cite{thomas_electrocardiogram-based_2005}. The CPC categorises sleep into: Deep Sleep (DS, corresponding to NREM N3 with parasympathetic dominance), REM Sleep (rapid eye movement with variable autonomic activity), and Rapid Sleep (RS, transitional states with intermediate autonomic characteristics). Sleep staging data were available for all 38 patients.

\subsection{Feature Extraction}
A total of 105 features were extracted across multiple domains:
\begin{enumerate}
    \item \textbf{Demographics (3 features):} Age, height, and weight.
    \item \textbf{Clinical Measurements (20 features):} The predictive models integrated a comprehensive set of clinical biomarkers across multiple physiological domains. These included hemodynamic measures such as systolic blood pressure (SBP) and diastolic blood pressure (DBP); a complete lipid panel comprising triglycerides (TG), high-density lipoprotein (HDL), and low-density lipoprotein (LDL); and indicators of renal function, specifically blood urea nitrogen (BUN), uric acid (UA), and the urinary albumin-to-creatinine ratio (UACR). Liver function was assessed using enzymes including alanine aminotransferase (ALT), aspartate aminotransferase (AST), AST/ALT ratio, and gamma-glutamyl transferase (GGT). Additionally, the feature set incorporated the inflammatory marker C-reactive protein (CRP) and key hematological parameters: white blood cell count (WBC), hemoglobin (Hb), platelets (PLT), and neutrophil percentage (NEUT\%). Finally, the status of lower extremity atherosclerosis (LEA) was included to account for vascular complications. A detailed description of all the 105 extracted features is provided in Supplementary Materials Table S1.
    \item \textbf{ECG Morphology (24 features):} Signal statistics (mean, standard deviation, range, signal-to-noise ratio [SNR] estimate) computed separately for 24-hour, sleep-period, and daytime recordings.
    \item \textbf{HRV Features (33 features):} Time-domain metrics per Task Force guidelines \cite{noauthor_heart_1996} computed for each sleep stage: mean RR interval, SDNN, RMSSD, pNN50, RR range, coefficient of variation, and recording duration.
    \item \textbf{Age-Normalized HRV (3 features):} Mean RR intervals normalized by age factor for each sleep stage.
    \item \textbf{Sleep Quality (22 features):} Pittsburgh Sleep Quality Index (PSQI) components including global score and seven sub-scores \cite{buysse_pittsburgh_1989} (11 features), and CPC-derived metrics including total/stage-specific sleep times, sleep stage proportions, and apnea-hypopnea index (11 features). 
\end{enumerate}

\subsection{Age Normalization Methodology}
The proposed age normalization applied the following transformation to HRV mean RR features:

\begin{equation}
    \mathrm{HRV}_{\mathrm{normalized}} = \frac{\mathrm{HRV}_{\mathrm{raw}}}{\left(\frac{\mathrm{age}}{65}+0.1\right)} 
\end{equation}

where 65 is a clinically significant age threshold taken from Umetani et al. \cite{umetani_twenty-four_1998}, who found a large age-related reduction in HRV, and 0.1 is included to ensure that the parameters remain numerically stable by avoiding division by numbers close to zero in younger patients. To verify that the negative result for the normalization of age is not simply an artefact of these specific choices of parameters, a sensitivity analysis was conducted over 20 different parameter settings (five different age thresholds: 55, 60, 65, 70, and 75 years, and four different stability constants: 0.05, 0.1, 0.15, and 0.2), as shown in Supplementary Materials Table S2-S3. In addition, five alternative age-adjustment strategies were evaluated to determine whether any fundamentally different approach to age correction improved estimation performance. To evaluate these approaches systematically, we compared six methods as described in Table~\ref{tab:age_adjustment_methods}.

\begin{table}[htbp]
\centering
\caption{Age Adjustment Methods (6 methods)}
\label{tab:age_adjustment_methods}
\begin{tabular}{ll}
\toprule
\textbf{Method} & \textbf{Approach} \\
\midrule
No adjustment (baseline) & Baseline \\
Proposed $\left(\frac{\mathrm{HRV}}{\frac{\mathrm{age}}{65}+0.1}\right)$ & Normalisation formula \\
Residualisation & Regress age out of HRV via linear regression \\
Age-Bin Z-Score & Z-score within age quartiles \\
Polynomial interaction & $\mathrm{Age}^2 + \mathrm{HRV} \times \mathrm{age}$ interaction terms \\
Simple division $\left(\frac{\mathrm{HRV}}{\mathrm{age}}\right)$ & Divide HRV by age directly \\
\bottomrule
\end{tabular}
\end{table}

\subsection{Feature Selection}

Feature selection employed SelectKBest with univariate F-regression, retaining up to 15 features by F-statistics within each CV fold. A separate false discovery rate (FDR)-corrected analysis (Benjamini-Hochberg procedure \cite{benjamini_controlling_1995}, threshold $p < 0.1$) was performed on the full dataset for descriptive reporting of feature-target associations only and was not used for model training.

Critically, both feature selection (SelectKBest with F-regression, $k = 15$) and standardisation (z-scoring) were performed strictly within each LOSO fold: fitted on training subjects only and applied to the held-out test subject using training-derived parameters. This prevents any information from the test subject influencing the feature selection or scaling, ensuring unbiased performance estimates \cite{varoquaux_cross-validation_2018}. Feature selection stability across folds was tracked to assess robustness (as described in detail in Section 4 of the Supplementary Materials); features selected in $>$80\% of folds were designated ``stable'', 50--80\% ``moderate'', and $<$50\% ``unstable''.

\subsection{Machine Learning Models}
We evaluated 20 algorithms spanning multiple paradigms:

\begin{enumerate}
    \item \textbf{Naive Baselines (2):} Mean predictor and median predictor, establishing minimum performance benchmarks.
    \item \textbf{Linear Models (7):} Ordinary least squares, Ridge ($\alpha=0.1, 1.0$), Lasso ($\alpha=0.1$), ElasticNet ($\alpha=0.1$, L1 ratio=0.5), Bayesian Ridge ($\alpha_1=\alpha_2=\lambda_1=\lambda_2=10^{-6}$), Huber regressor.
    \item \textbf{Tree Ensembles (4):} Random Forest, Extra Trees, Gradient Boosting (100 estimators, max depth 3-5, min samples leaf 3), AdaBoost.
    \item \textbf{Support Vector Machines (3):} Support vector regression (SVR) with radial basis function (RBF), linear, and polynomial (degree 2) kernels, C=1.0.
    \item \textbf{Neural Networks (4):} Multilayer perceptrons (MLP) with architectures (32), (64,32), and (128,64,32) hidden units with ReLU activation, plus an additional (64,32) variant with tanh activation; Adam optimizer; $\alpha=0.01$ regularization; early stopping.
\end{enumerate}

Minimally configured hyperparameters were used for all the models without nested CV tuning. This approach was chosen deliberately: (1) small sample sizes ($n = 29$--38) preclude reliable hyperparameter optimisation, (2) it provides a conservative baseline reflecting typical research practice, and (3) it enables fair comparison across model families under identical conditions. Scikit-learn defaults are designed to achieve a reasonable performance across problems \cite{pedregosa_scikit-learn_2018}. Therefore, neural network results should be interpreted as performance under these specific constraints rather than as a general assessment of neural architectures for physiological prediction tasks.

\subsection{Validation Strategy}

LOSO CV was implemented using Scikit-learn's LeaveOneGroupOut \cite{pedregosa_scikit-learn_2018}, with each subject constituting a separate test fold. This approach ensures complete independence between the training and test data, preventing information leakage that can occur with random k-fold splits in physiological datasets \cite{varoquaux_cross-validation_2018, saeb_need_2017}. For HbA1c, this yielded 29 folds; for FBG, 38 folds.

\subsection{Statistical Analysis}
The performance metrics included the coefficient of determination ($R^2$), mean absolute error (MAE), and Pearson correlation coefficient, along with significance testing. The statistical significance of model performance versus chance was assessed via permutation testing ($n = 500$ permutations) \cite{ojala_permutation_2010}. Uncertainty quantification employs bootstrap resampling ($n = 500$ iterations) to generate 95\% confidence intervals \cite{efron_introduction_1994}. Residual analysis assessed normality (using the Shapiro-Wilk test), bias (where the mean residual was significantly different from zero), and heteroscedasticity (the correlation between predictions and absolute residuals).

\subsection{Ablation Study Design}
Systematic ablation analysis quantified individual component contributions through 13 configurations: Full Model (all features), No Age Normalisation, Only Age-Normalised + Demographics, No Sleep-Stage HRV, HRV Only, ECG Only, Clinical Only, No ECG, No Clinical, Only Deep Sleep HRV, Only REM HRV, Only Rapid Sleep HRV, and Demographics Only. The Bayesian Ridge was used as the ablation model because its linear coefficient structure provides transparent attribution of performance changes to specific feature domains, whereas the inherent robustness of tree-based ensembles to irrelevant features attenuates the observable impact of domain removal.

\section{Results}
\label{subsec:results_discussion}

\subsection{Baseline Model Comparison}
\begin{figure}
    \centering
    \includegraphics[width=\linewidth]{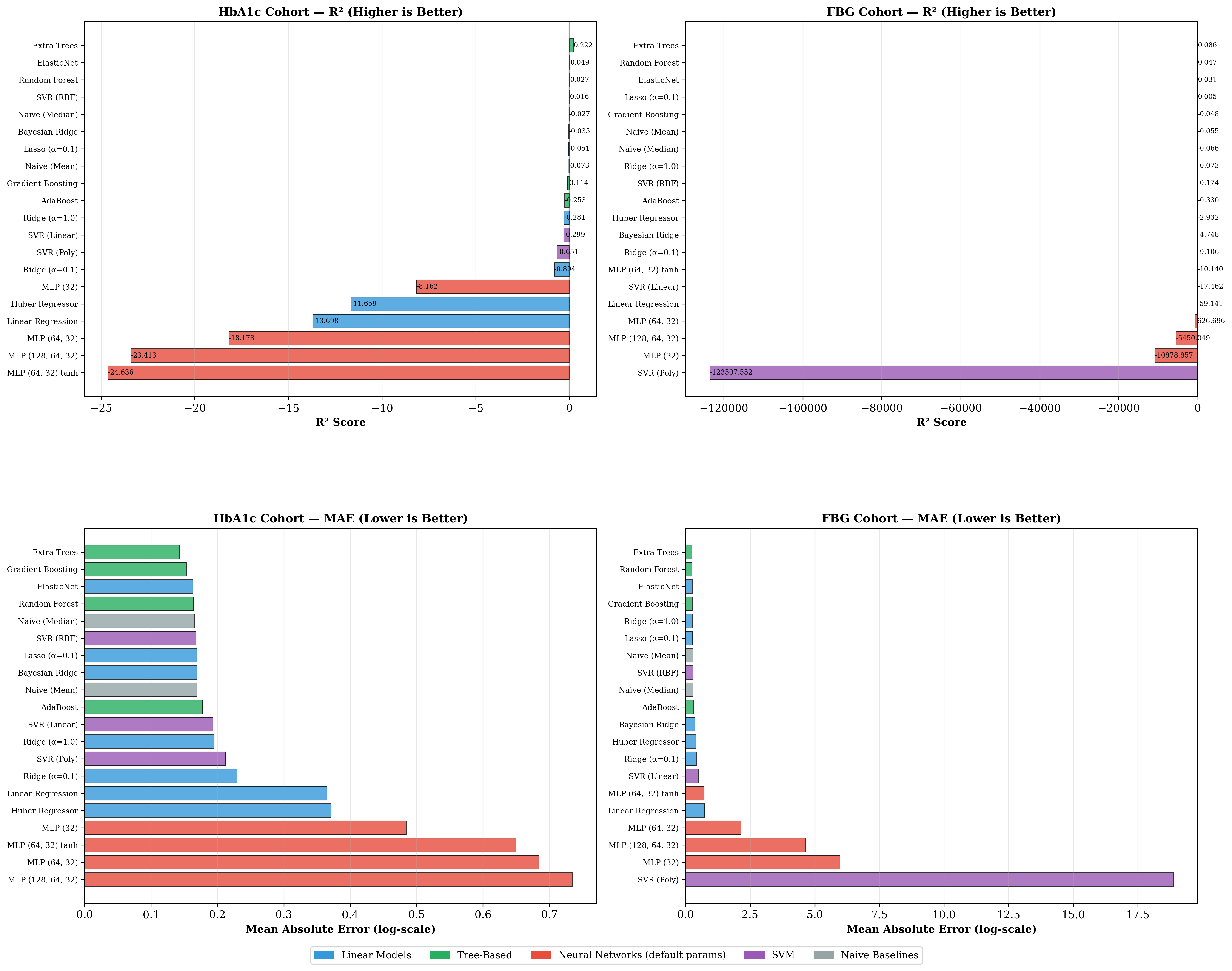} 
    \caption{Baseline model comparison under Leave-One-Subject-Out validation with within-fold feature selection and standardisation. Upper panels show $R^2$ scores (higher is better) for HbA1c (left) and FBG (right) cohorts. Lower panels show Mean Absolute Error on log-transformed targets (lower is better). Tree-based ensembles (green) maintained modest positive $R^2$ values, while neural networks with minimally configured hyperparameters (red) exhibited severe overfitting with $R^2$ values as low as $-24.6$ (HbA1c) and $-10{,}879$ (FBG).}
    \label{fig:feature_importance}
\end{figure}

The performance metrics for the models, using leave-one-subject-out validation with strict within-fold feature selection and standardization, are summarized in Table~\ref{tab:regression_performance} and graphically compared in Figure~\ref{fig:feature_importance}. For the HbA1c cohorts, the Extra Trees ensemble method had the highest $R^2$ of 0.222, followed by ElasticNet with $R^2 = 0.049$ and Random Forest with $R^2 = 0.027$. For FBG, the Extra Trees ensemble method performed best ($R^2 = 0.086$), followed by Random Forest ($R^2 = 0.047$) and ElasticNet ($R^2 = 0.031$). In all cases, the best-performing tree-based methods (Extra Trees, Random Forest) outperformed the naive models and linear regression. The Bayesian Ridge collapsed to $R^2 = -0.035$ for HbA1c and $R^2 = -4.748$ for FBG with strict within-fold preprocessing, demonstrating the vulnerability of the linear models to feature selection instability in small-sample settings. Neural networks with minimally configured hyperparameters performed poorly on both cohorts, achieving $R^2$ values ranging from $-8.2$ to $-24.6$ for HbA1c and from $-10.14$ to $-10{,}879$ for FBG. The strongly negative $R^2$ values, especially the ones that are in the thousands for FBG, suggest numerical instability, not a legitimate comparison. When the model overfits the noise present within the 28 to 37 training subjects and makes predictions on the single held-out subject, the squared error can become much greater than the total variance $(SS_{\mathrm{res}} \gg SS_{\mathrm{tot}})$, causing the $R^2$ values to become arbitrarily large and negative, as $R^2 = 1 - \frac{SS_{\mathrm{res}}}{SS_{\mathrm{tot}}}$. These results confirm that, under the given constraints ($n = 29$-38 subjects, 15 selected features, no architecture search), neural networks are suboptimal compared to regularized models and tree-based ensemble methods. It is likely that more extensive datasets ($n > 200$) and carefully crafted architectures can lead to competitive or superior performance.

From a clinical interpretability standpoint, back-transforming predictions to the original scale yielded a mean absolute error of 1.18 percentage points for HbA1c (mean absolute percentage error [MAPE]: 14.6\%) and 2.27 mmol/L (41 mg/dL) for FBG (MAPE: 24.9\%), as detailed in Supplementary Materials Table S4. For context, the log-scale MAE values of 0.142 (HbA1c) and 0.247 (FBG) correspond to approximate multiplicative error factors of $1.15\times$ and $1.28\times$ respectively, indicating that predictions deviate by approximately 15\% and 28\% from the actual values on average.

\begin{table}[H]
    \centering
    \setlength{\tabcolsep}{4pt}
    \caption{Regression performance of candidate models for HbA1c and FBG estimation using HRV-derived features}
    \label{tab:regression_performance}
    \begin{tabular}{lcccccc}
        \toprule
        & \multicolumn{3}{c}{\textbf{HbA1c}} & \multicolumn{3}{c}{\textbf{FBG}} \\
        \cmidrule(lr){2-4} \cmidrule(lr){5-7}
        \textbf{Model} & \textbf{R\textsuperscript{2}} & \textbf{MAE} & \textbf{Pearson r (p-value)} & \textbf{R\textsuperscript{2}} & \textbf{MAE} & \textbf{Pearson r (p-value)} \\
        \midrule
        Extra Trees* & 0.222 & 0.142 & 0.476 (0.009) & 0.086 & 0.247 & 0.344 (0.034) \\
        ElasticNet & 0.049 & 0.163 & 0.240 (0.210) & 0.031 & 0.261 & 0.235 (0.156) \\
        Random Forest & 0.027 & 0.164 & 0.269 (0.158) & 0.047 & 0.255 & 0.329 (0.043) \\
        SVR (RBF) & 0.016 & 0.168 & 0.236 (0.217) & $-0.174$ & 0.288 & 0.189 (0.255) \\
        Bayesian Ridge & $-0.035$ & 0.169 & 0.219 (0.254) & $-4.748$ & 0.360 & 0.255 (0.123) \\
        Gradient Boosting & $-0.114$ & 0.153 & 0.265 (0.165) & $-0.048$ & 0.263 & 0.333 (0.041) \\
        Naïve (Mean) & $-0.073$ & 0.169 & -- & $-0.055$ & 0.287 & -- \\
        MLP (64, 32) & $-18.178$ & 0.684 & 0.128 (0.507) & $-626.7$ & 2.150 & $-0.183$ (0.271) \\
        \bottomrule
    \end{tabular}
    \vspace{2pt}
    \begin{minipage}{0.95\textwidth}
    \footnotesize
    *Best performing model for the cohort. The MAE was computed on log-transformed targets. All models were evaluated using strict within-fold feature selection and standardisation.
    \end{minipage}
\end{table}

\subsection{Age Normalisation: Negative Finding}
Contrary to the primary hypothesis, age normalization offered neither benefit nor improvement under the stringent within-fold validation shown in Table~\ref{tab:age_adjustment_comparison}. Compared to the Bayesian Ridge model as a baseline for comparison, HbA1c $R^2$ changed from $-0.035$ without normalization to $-0.040$ with the new normalization strategy, and FBG $R^2$ changed from $-4.748$ to $-4.951$. Each of the six age-adjustment strategies, including the new normalization strategy, residualization, age-bin z-scores, polynomial interactions, and simple division, offered no clinically meaningful improvement over the unadjusted model, with most methods performing comparably or worse in both cohorts. A full sensitivity analysis was conducted over 20 different parameter settings (five age thresholds: 55, 60, 65, 70, and 75 years; four stability constants $\varepsilon$: 0.05, 0.10, 0.15, and 0.20) to confirm the validity of the negative results (as detailed in Section 2 of the Supplementary Materials). For HbA1c, all normalized $R^2$ values were between $-0.071$ and $-0.035$, compared to a baseline of $-0.035$; only one setting (threshold of 65 years, $\varepsilon = 0.15$) offered a trivial improvement ($\Delta R^2 = +0.0001$). For the FBG, all 20 settings worsened the performance. The age threshold of 65 years was chosen in line with the findings of Umetani et al. \cite{umetani_twenty-four_1998}, who found that the rate of HRV reduction was maximal beyond this age, although no parameter settings offered any improvement.

\begin{table}[htbp]
\caption{Age adjustment method comparison (Bayesian Ridge model under LOSO validation with within-fold preprocessing)}
\label{tab:age_adjustment_comparison}
\centering
\begin{tabular}{lcc}
\toprule
\textbf{Method} & \textbf{HbA1c $R^2$} & \textbf{FBG $R^2$} \\
\midrule
No adjustment (baseline) & $-0.035$ & $-4.748$ \\
Proposed $\left(\frac{\mathrm{HRV}}{\frac{\mathrm{age}}{65}+0.1}\right)$ & $-0.040$ & $-4.951$ \\
Residualisation & $-0.035$ & $-4.748$ \\
Age-Bin Z-Score & $-0.068$ & $-4.748$ \\
Polynomial interaction & $-0.099$ & $-3.941$ \\
Simple division $\left(\frac{\mathrm{HRV}}{\mathrm{age}}\right)$ & $-0.082$ & $-4.953$ \\
\bottomrule
\end{tabular}
\end{table}

\subsection{Feature Domain Contributions}

\begin{figure}
    \centering
    \includegraphics[width=\linewidth]{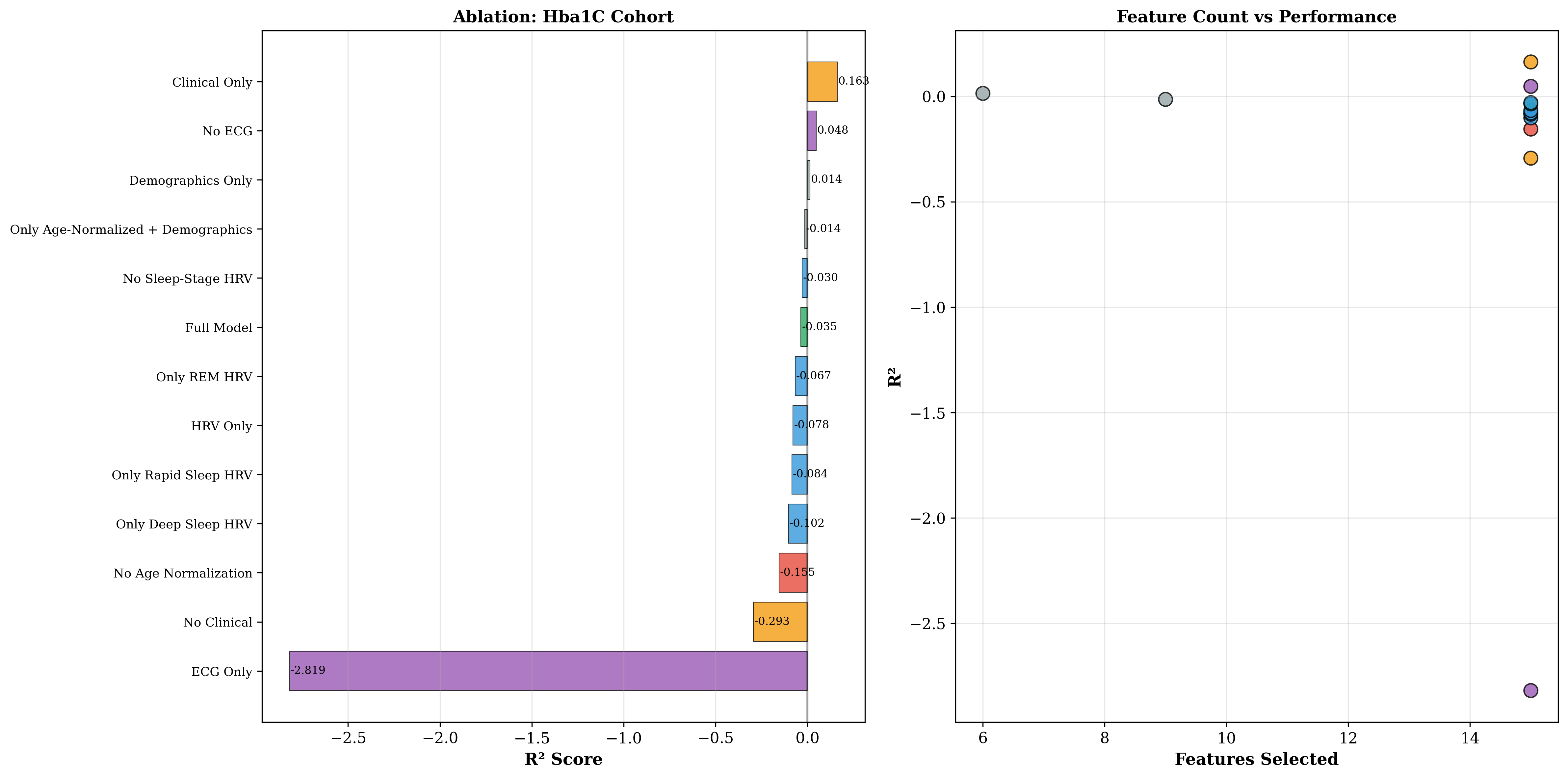}
    \caption{Ablation study results of HbA1c cohort: $R^2$ by feature configuration (left) and feature count vs. performance (right).}
    \label{fig:ablation_study_results_1}
\end{figure}

\begin{figure}
    \centering
    \includegraphics[width=\linewidth]{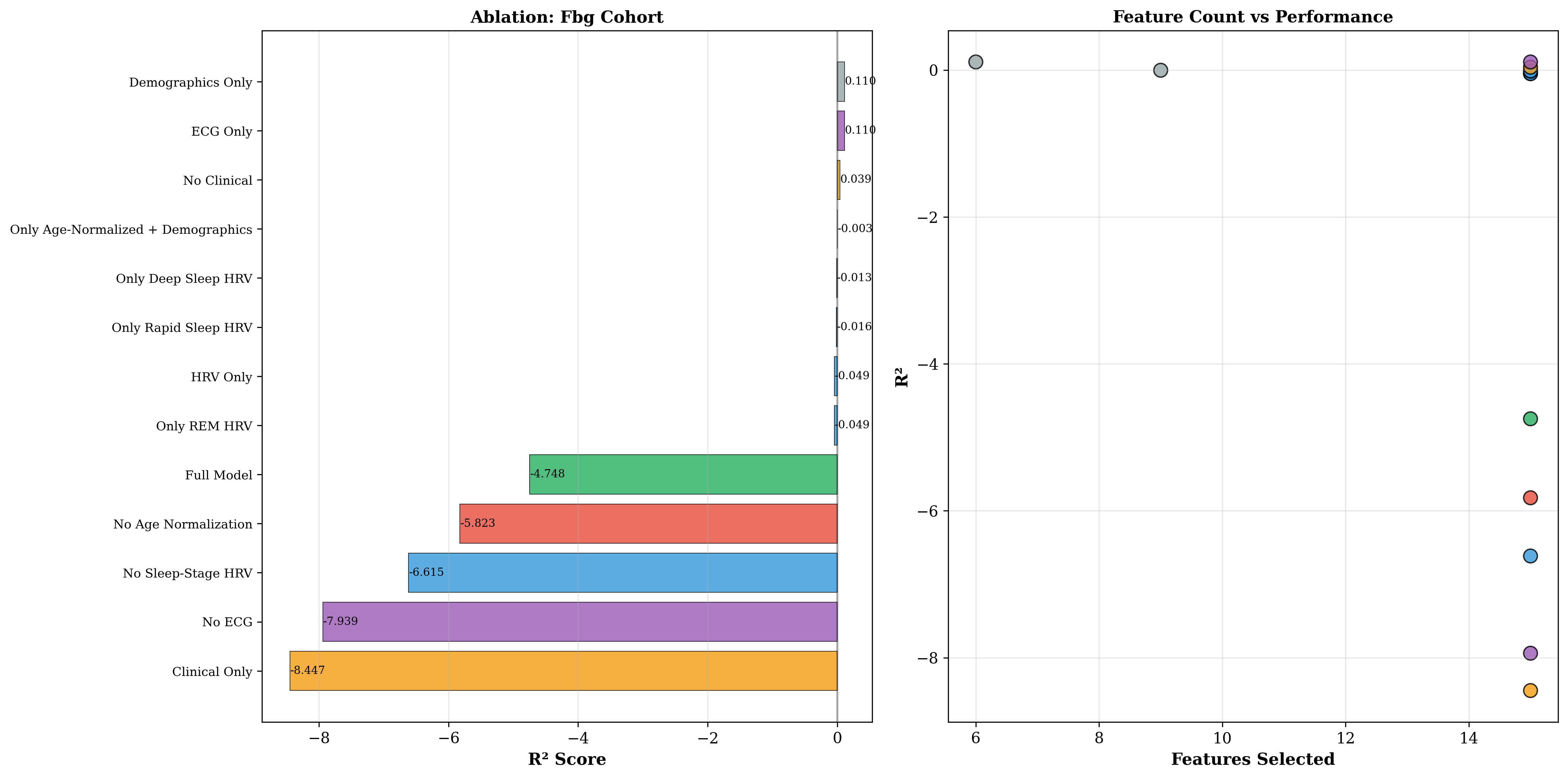}
    \caption{Ablation study results of FBG cohort: $R^2$ by feature configuration (left) and feature count vs. performance (right).}
    \label{fig:ablation_study_results_2}
\end{figure}

Analysis of ablation, performed under strict within-fold validation, showed cohort-specific trends, as presented in Table~\ref{tab:ablation_configurations}, Figures~\ref{fig:ablation_study_results_1}--\ref{fig:ablation_study_results_2}. For HbA1c, the model with clinical features alone reached $R^2 = 0.163$, the best-performing ablation configuration with a positive $R^2$ value, outperforming the Full Model ($R^2 = -0.035$). This result indicates that adding ECG and HRV features to clinical data introduces noise that hurts the performance under strict within-fold feature selection on small samples. Models with HRV features alone performed poorly ($R^2 = -0.078$), and ECG morphology alone showed strong signs of overfitting ($R^2 = -2.819$). Demographics alone reached $R^2 = 0.014$.

For FBG, the demographics-only and ECG-only models reached $R^2 = 0.110$, whereas most models with larger feature sets had strongly negative $R^2$ values (Full Model: $R^2 = -4.748$; Clinical Only: $R^2 = -8.447$), indicating that small-sample estimates are prone to feature selection instability when the feature set size is large compared to the number of subjects. This difference in cohort-specific trends mirrors the different underlying physiological concepts of the targets: HbA1c is a measure of long-term metabolism, well-integrated by clinical data, while FBG is a measure of acute autonomic and metabolic interactions, where simpler feature sets (demographics and ECG morphology) provide more stable estimates.

\begin{table}[htbp]
\caption{Ablation study: Feature domain contributions (Bayesian Ridge, LOSO validation)}
\label{tab:ablation_configurations}
\centering
\begin{tabular}{lcc}
\toprule
\textbf{Configuration} & \textbf{HbA1c $R^2$} & \textbf{FBG $R^2$} \\
\midrule
Full Model              & $-0.035$  & $-4.748$ \\
Clinical Only           & $\mathbf{0.163}^{*}$ & $-8.447$ \\
No ECG                  & $0.048$   & $-7.939$ \\
Demographics Only       & $0.014$   & $\mathbf{0.110}^{*}$ \\
ECG Only                & $-2.819$  & $\mathbf{0.110}^{*}$ \\
HRV Only                & $-0.078$  & $-0.049$ \\
No Clinical             & $-0.293$  & $0.039$ \\
No Age Normalisation    & $-0.155$  & $-5.823$ \\
\bottomrule
\end{tabular}

\vspace{2pt}
\begin{minipage}{0.95\textwidth}
\footnotesize
$^{*}$Best configuration for the cohort. Bayesian Ridge with within-fold feature selection and standardisation.
\end{minipage}
\end{table}

\begin{figure}
    \centering
    \includegraphics[width=\linewidth]{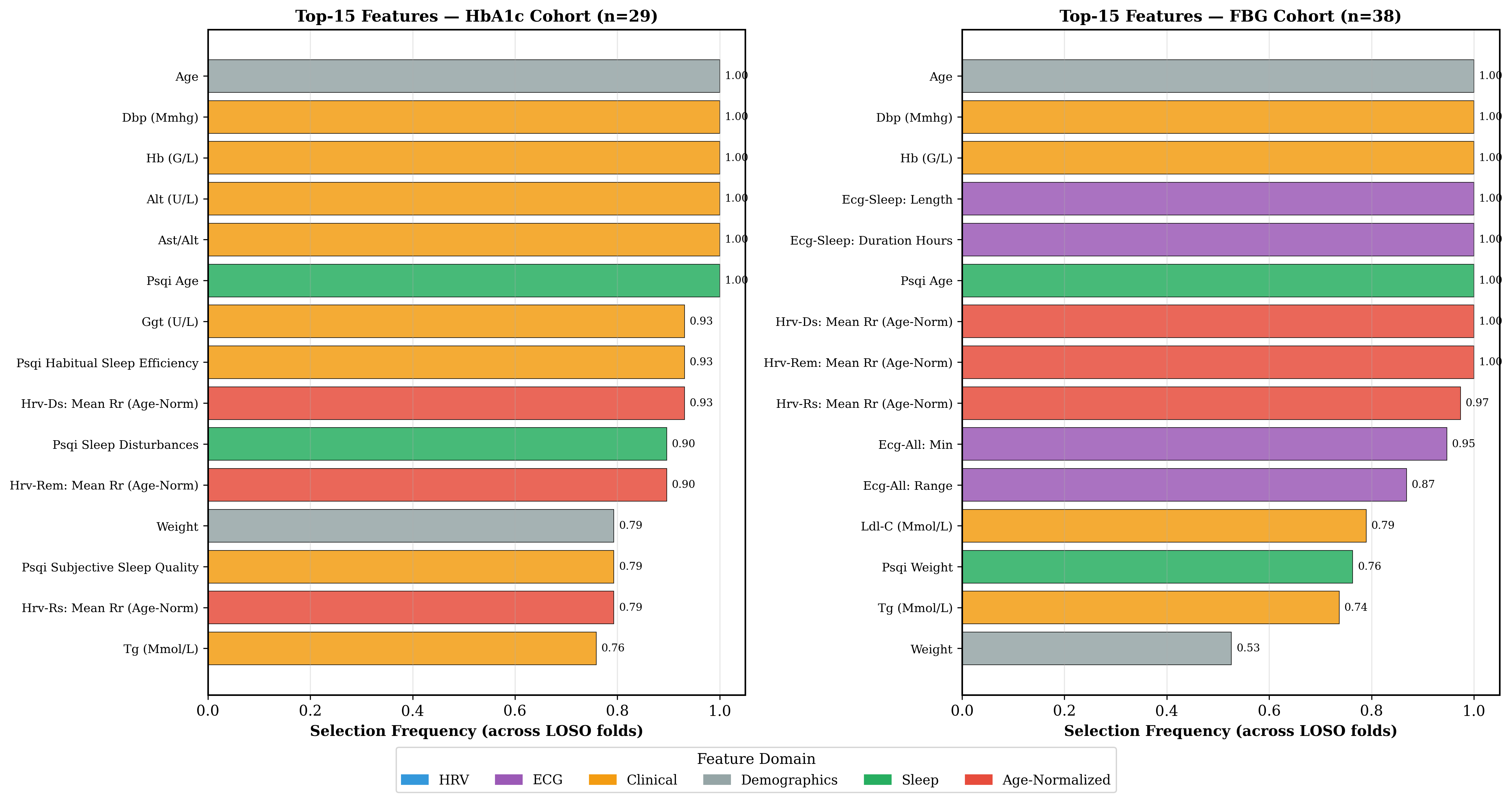}
    \caption{Domain-specific feature reliance for long-term (HbA1c) versus short-term (FBG) glycemic estimation. Features are ranked by their stability (frequency of selection) across LOSO folds.}
    \label{fig:domain_specific_feature}
\end{figure}

Feature importance analysis, as shown in Figure~\ref{fig:domain_specific_feature}, reveals the specific features that drive the estimation performance within each domain. For HbA1c, the most stably selected features (100\% of LOSO folds) were age, diastolic blood pressure, haemoglobin, ALT, AST/ALT ratio, and PSQI age score, which were predominantly clinical and demographic features. For FBG, age, diastolic blood pressure, hemoglobin, ECG sleep duration, PSQI age, and age-normalised HRV mean RR intervals during all three sleep stages were selected in $\geq$97\% of folds, showing greater ECG/HRV involvement consistent with the ablation findings.

\subsection{Statistical Validation}

\begin{figure}
    \centering
    \includegraphics[width=\linewidth]{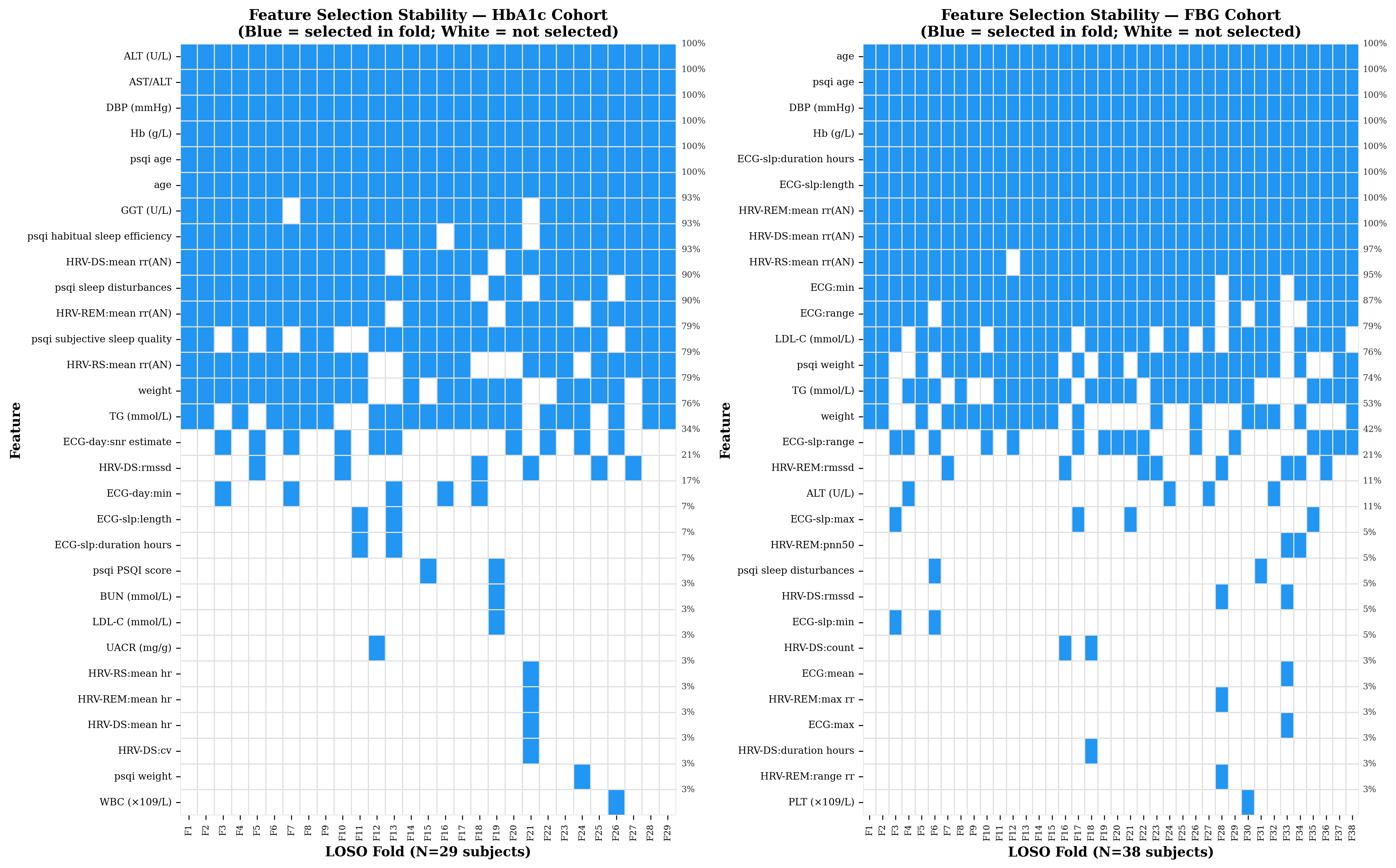}
    \caption{Feature selection stability across CV folds. Binary heatmaps illustrating the consistency of feature selection during Leave-One-Subject-Out (LOSO) validation for the (A) HbA1c and (B) FBG cohorts. Blue cells indicate features selected by the SelectKBest algorithm within a specific fold; white cells indicate non-selection.}
    \label{fig:feature_selection_stability}
\end{figure}

The feature selection stability analysis as shown in Figure~\ref{fig:feature_selection_stability}, provides direct evidence of within-fold feature selection. The binary heatmap visualises which features were selected in each LOSO fold, with clear variation across folds confirming that SelectKBest was fitted independently within each fold rather than on the full dataset. Ten features were selected in $>$80\% of folds for the HbA1c cohort (designated as `stable'), including clinical markers, such as AST/ALT and hemoglobin. In contrast, the FBG cohort showed stable selection of ECG-derived features, such as sleep duration and age-normalised HRV, reflecting the domain-specific drivers identified in the importance analysis.

The permutation test with $n = 500$ iterations showed that the performance of the best model (Extra Trees) was significantly better than chance for both the groups ($p = 0.002$). The obtained $R^2$ values were 0.222 for HbA1c and 0.086 for FBG, both above the 99.8th percentile of the permutation distribution. The bootstrap 95\% confidence intervals ($n = 500$, subject-level resampling) were as follows: HbA1c $R^2$ in [0.13, 0.82] (median 0.550) and FBG $R^2$ in [0.10, 0.72] (median 0.472). These intervals did not contain zero for either group, suggesting that there is statistical evidence that the observed predictive associations are not likely to be due to chance alone. Notably, the bootstrap confidence intervals are substantially wider than the point estimates, which is expected given the small sample sizes. Cross-validation and bootstrap estimates are known to exhibit inherently large error bars when the number of samples is small \cite{varoquaux_cross-validation_2018}; with only 29--38 individuals in our cohorts, each bootstrap resample produces a different composition of subjects, and the inclusion or exclusion of even a single influential subject can substantially shift the $R^2$ estimate. This instability is a known statistical property of small-sample cross-validation rather than a flaw in the modeling approach. Nevertheless, due to the very small sample sizes ($n = 29$ for HbA1c, $n = 38$ for FBG), the fact that this study was conducted at a single centre with a male-only population, and the fact that this study used spot glycemic measurements rather than CGM, these findings should be considered as preliminary evidence of an HRV-glycemic association that requires validation in larger and more diverse populations.

\begin{figure}
    \centering
    \includegraphics[width=\linewidth]{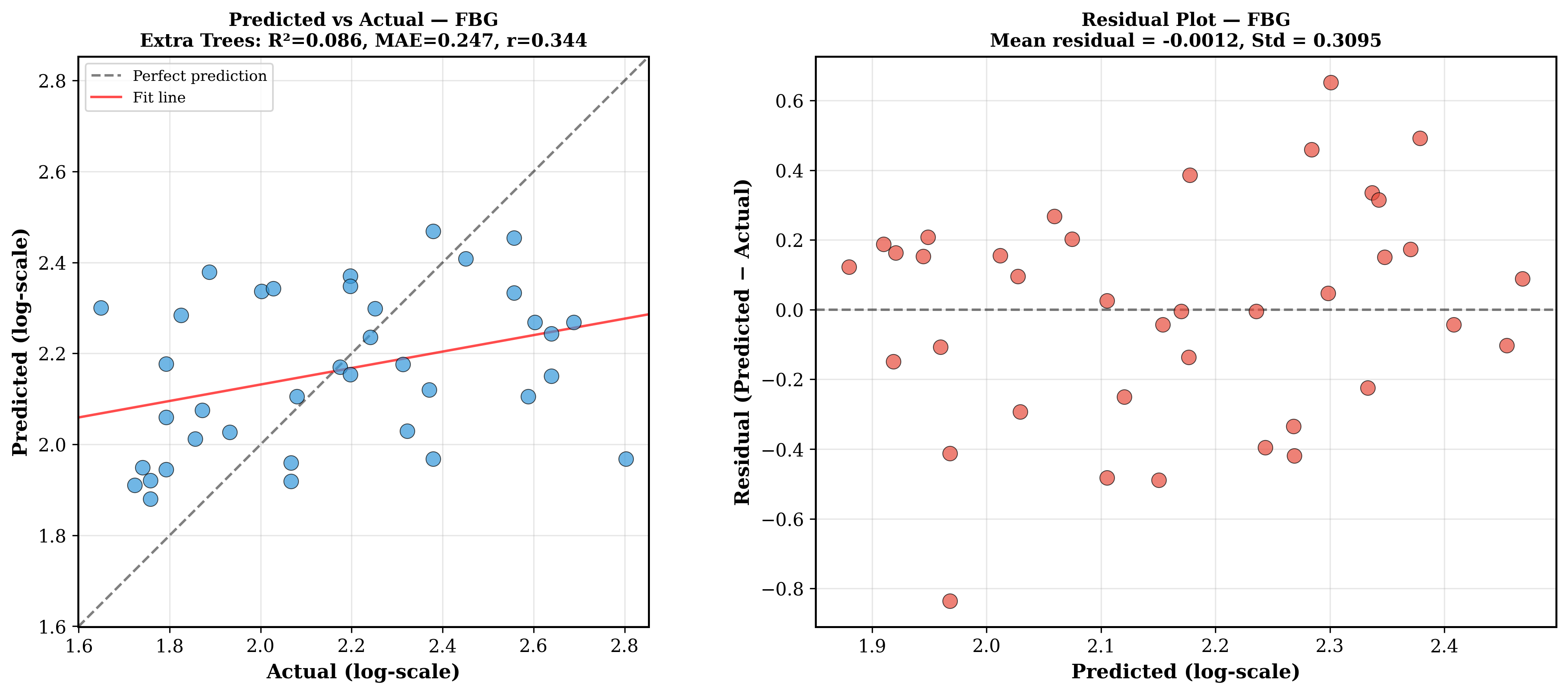}
    \caption{Prediction performance for FBG cohort (n = 38), Extra Trees model. (Left) Predicted vs. actual log-transformed FBG ($R^2 = 0.086$, MAE = 0.247, r = 0.344). Dashed line indicates perfect prediction; red line shows linear fit. (Right) Residual distribution showing near-zero mean bias and no significant heteroscedasticity.}
    \label{fig:prediction_performance_fbg}
\end{figure}

\begin{figure}
    \centering
    \includegraphics[width=\linewidth]{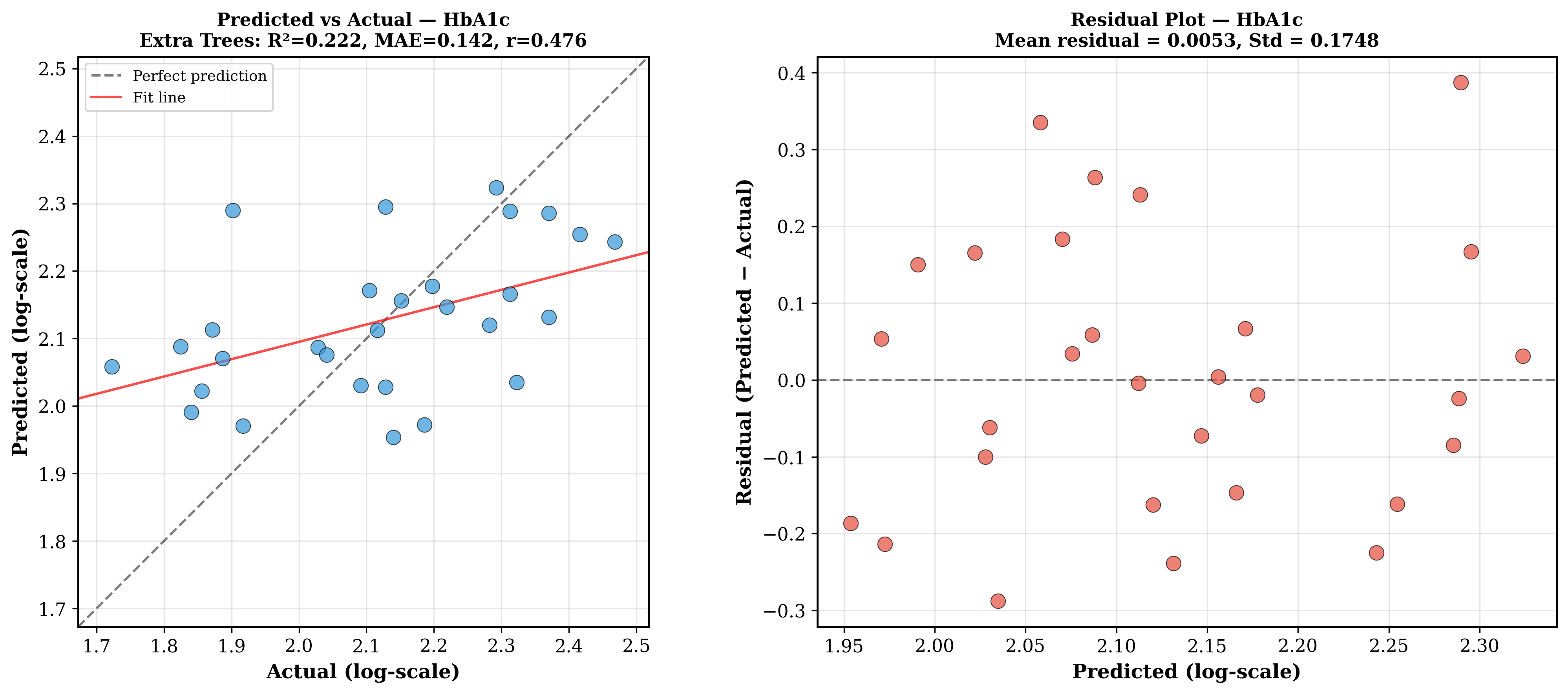}
    \caption{Prediction performance for HbA1c cohort (n = 29), Extra Trees model. (Left) Predicted vs. actual log-transformed HbA1c values ($R^2 = 0.222$, MAE = 0.142, r = 0.476). Dashed line indicates perfect prediction. (Right) Residual analysis showing no systematic bias ($\text{mean residual} \approx 0$) and approximately normal distribution.}
    \label{fig:prediction_performance_hba1c}
\end{figure}

The residual tests for the best model (Extra Trees) did not show any significant estimation bias for either group, because the mean residual was not significantly different from zero. For HbA1c, the residuals were normally distributed (Shapiro-Wilk $p = 0.61$) with no significant heteroscedasticity ($r = -0.09$, $p = 0.63$). For FBG, the residuals were also normally distributed (Shapiro-Wilk $p = 0.80$) with no significant heteroscedasticity ($r = 0.01$, $p = 0.96$), which indicates that the prediction errors were well-behaved over the range of the observed values. The results of the learning curve test showed that the performance for HbA1c was saturated at the present sample size, whereas FBG estimation could be improved with more subjects. The accuracy of the prediction and distribution of residuals are shown in Figures~\ref{fig:prediction_performance_fbg}--\ref{fig:prediction_performance_hba1c}.

\section{Discussion}

\subsection{Principal Findings}

The main result of this research is that strict validation procedures, especially CV hygiene, lead to results that are significantly different from those obtained by traditional approaches in the analysis of HRV-based glycemic status estimation. Two main results were obtained. First, there is a strong differential effect of CV hygiene on model families. When the effect of within-fold feature selection was properly accounted for by applying SelectKBest and StandardScaler to the training data, the effect was highly dependent on the model architecture. Tree-based models (Extra Trees $R^2 = 0.222$, Random Forest $R^2 = 0.027$ for HbA1c) showed only a slight positive effect, whereas the Bayesian Ridge collapsed to $R^2 = -0.035$ for HbA1c and $-4.748$ for FBG, demonstrating the vulnerability of linear models to feature selection instability when preprocessing is not confined within CV folds. This stark contrast between model families illustrates the magnitude of bias that improper CV practices can introduce and emphasizes that subject-independent validation with within-fold preprocessing should represent the bare minimum in small-sample physiological estimation research. The differential performance is due to the implicit feature selection inherent in tree ensembles through split-based importance and robustness to irrelevant features, as opposed to linear models that propagate noise from improperly chosen features directly into coefficient estimates. These results have the following implications for small-sample research in physiology. In cases where feature selection is necessary within CV folds, tree-based models are preferred. Second, the primary hypothesis that age normalization would improve the accuracy of the estimation is clearly rejected. None of the variants of normalization provided significant improvement under the stringent subject-independent validation procedure for either group. Sensitivity analyses for 20 parameter settings (five age thresholds $\times$ four stability constants) supported this negative finding. For HbA1c, all values of normalized $R^2$ were between $-0.071$ and $-0.035$ compared to a baseline of $-0.035$; only one setting (threshold 65, $\varepsilon = 0.15$) showed a negligible improvement ($\Delta R^2 = +0.0001$). For the FBG, all 20 settings worsened the performance. The previous claims of the benefits of age normalization \cite{stojmenski_age_2023} might be due to various methodological artefacts such as mixed glucose targets, random CV allowing information transfer, or poor baseline comparisons. Our thorough investigation of six age-adjustment strategies offers strong evidence against simplistic age normalization for ECG-based glycemic estimation. 

From a clinical viewpoint, the best $R^2$ of 0.222 for HbA1c indicates that heart rate variability-derived variables explain approximately 22\% of the inter-subject variability in the log-transformed glycemic indices, which is not sufficient for individual clinical prediction but is in line with other single-modality, non-invasive estimation methods. For FBG, the best $R^2$ value of 0.086 explained less than 9\% of the variability. For a clinically viable non-invasive glycemic estimation, an $R^2$ value above 0.7 would be required, with errors confined to Clarke Error Grid zones A and B \cite{clarke_evaluating_1987}. These results show a statistically significant but clinically preliminary association between HRV and glycemic indices.

\subsection{Limitations}

There are a number of key limitations to the current study that must be considered when interpreting the results. First, although the bootstrap confidence intervals do not contain zero for either group, HbA1c $R^2$ [0.13, 0.82] (median 0.550) or FBG $R^2$ [0.10, 0.72] (median 0.472), the intervals are still wide, representing the considerable uncertainty associated with small-sample estimation. Such wide intervals are expected: Varoquaux \cite{varoquaux_cross-validation_2018} demonstrated that cross-validation error bars are inherently large for sample sizes below 100, as the performance estimate is dominated by sampling variability rather than model performance. In our cohorts, with only 29--38 subjects, $R^2$ is particularly sensitive to resampling because it involves a ratio of variance estimates that amplifies sampling fluctuations. The lower bounds are close to zero, especially for FBG, suggesting that although there is statistical support for the existence of a predictive relationship, the precision of the performance estimates remains limited. Second, the study was restricted to a male-only, single-ethnicity (Chinese) T2DM inpatient population, and thus cannot be generalized to females, other ethnic groups, type 1 diabetes, or outpatient populations. Sex differences in autonomic function, ethnic differences in HRV normal ranges, and inpatient settings may affect the relationship between HRV and glycemia. Third, the current study is restricted to spot glucose values (HbA1c and FBG) and thus cannot provide information on temporal relationships. Future studies should use simultaneous CGM and continuous ambulatory ECG to assess the temporal relationship between HRV and glucose levels. Fourth, the 28.3\% ECG exclusion rate may have been subject to selection bias, with patients having higher-quality recordings possibly differing systematically from those excluded from the analysis. Fifth, no nested hyperparameter optimization was performed. Although this provides conservative baselines, other models may rank differently when optimized. Taken together, these points provide a context in which our results may be considered preliminary evidence for an HRV-glycemic relationship, which should be validated in future multi-site, sex-balanced, ethnically diverse prospective studies with sufficient sample sizes ($n > 200$).

\subsection{Future Directions}

Future studies should employ simultaneous CGM and continuous ambulatory electrocardiography to investigate within-subject temporal relationships between HRV and glucose levels. Studies should use multi-site, sex-balanced, and ethnically diverse populations ($n > 200$) to assess generalisability and enable proper training of neural networks. The Clarke Error Grid analysis should replace $R^2$ as the primary measure of clinical relevance. To assess the clinical relevance of ECG-based glucose estimation in real-world settings, prospective outpatient studies should investigate whether HRV features can improve glycemic risk classification when integrated into wearable health platforms or serve as complementary inputs to CGM systems, potentially improving alarm specificity or sensor calibration intervals. Despite our rigorous assessment of six different age adjustment strategies, which produced negative results under rigorous within-fold validation, future studies with significantly larger and more diverse populations may wish to re-explore non-linear or data-driven approaches to age correction. Specifically, machine learning algorithms that can model complex interactions among age, HRV, and glycemia across broader age ranges may reveal patterns that are not accessible to simple parametric adjustments.

Recent work using CGM-derived dynamic features such as entropy rate and Poincaré plot indices to monitor glycemic state transitions \cite{montaser_novel_2024}, in combination with machine learning-based immunological risk stratification from standardized meal challenge CGM data \cite{montaser_enhancing_2025}, illustrates the general utility of physiologically informed, data-driven approaches to monitoring metabolic change. Such strategies may be used in conjunction with emerging non-invasive biomarkers, such as ECG- and HRV-based features, to provide comprehensive glycemic monitoring.

\section{Conclusions}

This reappraisal of methodological best practices clearly shows that CV hygiene is a fundamental determinant of which model families achieve success in glycemic status estimation from HRV data. Under stringent within-fold feature selection and standardization, tree-based ensembles showed limited positive results (Extra Trees: $R^2 = 0.222$ for HbA1c; $R^2 = 0.086$ for FBG), whereas linear models, including Bayesian Ridge, deteriorated to negative $R^2$ values ($-0.035$ for HbA1c; $-4.748$ for FBG), as expected owing to feature selection instability in small sample sizes. This result emphasizes that subject-independent validation with within-fold preprocessing should be the minimum for any physiological estimation study.

Age normalization did not provide any practical advantage under any of the 20 sensitivity settings and six age-adjustment methods for both groups. Clinical variables alone were the best estimators of HbA1c ($R^2 = 0.163$), whereas ECG morphology and demographics contributed the most to FBG estimation ($R^2 = 0.110$ each). Neural networks with minimally configured hyperparameters performed abysmally on these sample sizes (HbA1c $R^2$ range: $-8.2$ to $-24.6$; FBG $R^2$ range: $-10.1$ to $-10{,}879$), with the extreme negative values reflecting numerical instability inherent to LOSO evaluation on very small samples rather than a meaningful performance comparison; however, these results are more a reflection of the specific challenges of small tabular data and minimally configured hyperparameters than a general unsuitability of these models.

This study set the following minimum methodological requirements: (1) strict separation of fundamentally different glycemic endpoints; (2) subject-independent Leave-One-Subject-Out (LOSO) validation with within-fold feature selection and standardization; (3) thorough baseline comparisons across 20 algorithms; (4) explicit recognition of the limitations of cross-sectional spot measurements; and (5) proper reporting of uncertainty with bootstrap 95\% confidence intervals not containing zero for both groups (HbA1c $R^2$: [0.13, 0.82]; FBG $R^2$: [0.10, 0.72]), which provides statistical evidence for genuine predictive associations, while remaining wide due to the inherent instability of $R^2$ estimation in small samples. Before clinical use, these preliminary results must be validated in larger ($n > 200$) sex-balanced and ethnically representative cohorts with simultaneous CGM and ambulatory ECG recordings.

\vspace{0.2cm}
\noindent\textbf{Abbreviations:}
(ALT) Alanine aminotransferase,
(ANS) Autonomic nervous system,
(AST) Aspartate aminotransferase,
(BUN) Blood urea nitrogen,
(CAN) Cardiac autonomic neuropathy,
(CGM) Continuous glucose monitoring,
(CPC) Cardiopulmonary coupling,
(CRP) C-reactive protein,
(CV) Cross-validation,
(DS) Deep sleep,
(ECG) Electrocardiogram,
(FBG) Fasting blood glucose,
(GGT) Gamma-glutamyl transferase,
(HbA1c) Glycated hemoglobin,
(HDL) High-density lipoprotein,
(HF) High frequency,
(HRV) Heart rate variability,
(LDL) Low-density lipoprotein,
(LF) Low frequency,
(LOSO) Leave-one-subject-out,
(MAE) Mean absolute error,
(MAPE) Mean absolute percentage error,
(MLP) Multilayer perceptron,
(NREM) Non-rapid eye movement,
(pNN50) Percentage of successive RR intervals differing by more than 50 ms,
(PPG) Photoplethysmography,
(PRV) Pulse rate variability,
(PSQI) Pittsburgh Sleep Quality Index,
(RBF) Radial basis function,
(REM) Rapid eye movement,
(RMSSD) Root mean square of successive differences,
(RR) R-peak to R-peak interval,
($R^2$) Coefficient of determination,
(SDNN) Standard deviation of NN (normal-to-normal) intervals,
(SNR) Signal-to-noise ratio,
(SVR) Support vector regression,
(T2DM) Type 2 diabetes mellitus,
(WBC) White blood cell count.

\section*{Acknowledgements}
The authors acknowledge support from the Google Cloud Research Credits program under Award GCP19980904 and partial computing resources from Google's TPU Research Cloud (TRC), both of which provided critical infrastructure for this research.

\section*{Author Contributions} MBA conceptualized the study, designed the methodology, and curated the data. MBA and SIS conducted the formal analyses and investigations. MBA wrote the original draft and prepared the visualizations. SIS supervised the project, administered the study and contributed to resource acquisition. Both authors validated the results. All authors have read and approved the final manuscript.

\section*{Funding Source}
This study did not receive any specific grants from public, commercial, or not-for-profit funding agencies.

\section*{Data Availability} The dataset used and/or analysed during the current study is publicly available in the Mendeley Data repository, accessible at \url{https://data.mendeley.com/datasets/9c47vwvtss/4}. The processed data required to reproduce these findings, along with the code used for the analysis in this study, are available at \url{https://github.com/mdbasit897/HRV-Glycemic-Validation}.

\section*{Declarations}
\subsection*{Ethics Approval and Consent to Participate} This study involved a secondary analysis of a publicly available dataset from the Mendeley Data repository \cite{dataset_2023}. ``The original study was approved by the Institutional Review Board of the Suzhou Science \& Technology Town Hospital (No. IRB2019045), and all participants were informed of the experimental protocol and matters needing attention and then signed a written informed consent prior to participating in the study.'' No new human participants were recruited or directly involved in this study, and the data were de-identified; therefore, additional ethical approval was not required. 

\subsection*{Consent for Publication} Not Applicable.

\subsection*{Conflict of Interest Disclosure} The authors declare that they have no known competing financial interests or personal relationships that could influence the work reported in this study.

\section*{Supplementary Information}
Supplementary material associated with this article is available
\href{https://static-content.springer.com/esm/art\%3A10.1186\%2Fs12911-026-03455-8/MediaObjects/12911_2026_3455_MOESM1_ESM.docx}{online}.

\end{document}